\documentclass[fleqn,usenatbib]{mnras}

\usepackage{newtxtext,newtxmath}

\usepackage[T1]{fontenc}
\usepackage{ae,aecompl}

\usepackage{graphicx}	
\usepackage{amsmath}	
\usepackage{amssymb}	

\makeatletter
\newsavebox\myboxA
\newsavebox\myboxB
\newlength\mylenA

\newcommand*\xoverline[2][0.75]{%
    \sbox{\myboxA}{$\m@th#2$}%
    \setbox\myboxB\null
    \ht\myboxB=\ht\myboxA%
    \dp\myboxB=\dp\myboxA%
    \wd\myboxB=#1\wd\myboxA
    \sbox\myboxB{$\m@th\overline{\copy\myboxB}$}
    \setlength\mylenA{\the\wd\myboxA}
    \addtolength\mylenA{-\the\wd\myboxB}%
    \ifdim\wd\myboxB<\wd\myboxA%
       \rlap{\hskip 0.5\mylenA\usebox\myboxB}{\usebox\myboxA}%
    \else
        \hskip -0.5\mylenA\rlap{\usebox\myboxA}{\hskip 0.5\mylenA\usebox\myboxB}%
    \fi}
    
\title[On the use of first-order moment to measure $H_{\rm eff}$]{On the use of the first-order moment approach for 
measurements of $H_{\rm eff}$ from LSD profiles}

\author[J.C. Ram\' irez V\'elez]{
J.C. Ram\' irez V\'elez$^{1}$\thanks{E-mail: jramirez@astro.unam.mx}
\\
$^{1}$Instituto de Astronom\'ia - Universidad Nacional Aut\'onoma de M\'exico, 
           APO. Postal 877, 22860, Ensenada B.C.,  Mexico
}

\date{Accepted XXX. Received TY; in original form ZZZ}

\pubyear{2019}

\begin{document}
\label{firstpage}
\pagerange{\pageref{firstpage}--\pageref{lastpage}}
\maketitle
\newpage

\begin{abstract}
  The big majority of the reported measurements of the stellar magnetic fields that have analysed spectropolarimetric
  data have employed the least-square-deconvolution method (LSD) and the first-order moment approach.
  We present a series of numerical tests in which we review some important aspects
  of this technique. First, we show that the selection of the profile widths, i.e. integration range
  in the first-order moment equation, is independent of the accuracy of the magnetic measurements,
  meaning that for any arbitrary profile width it is always possible to properly determine the 
  longitudinal  magnetic field.  We also  study the interplay
  between the line depth limit adopted in the line mask and the normalisation values of the LSD profiles.
  We finally show that the rotation of the stars has to be considered to correctly infer the intensity of the 
  magnetic field, something that has been neglected up to now. We show that the latter
  consideration is crucial, and our test shows that the magnetic intensities differ by a factor close to 3 
  for a moderate fast rotator star with $vsini$ of 50 ${\rm km\,s^{-1}}$. Therefore, it is expected that
  in general the stellar magnetic fields reported for fast rotators are stronger than what was believed.  
  All the previous results shows that the first-order moment can be a very robust tool for measurements of magnetic fields,
  provided that the weak magnetic field approximation is secured. We also show that when the magnetic field regime
  breaks down, the use of the first-order moment method becomes uncertain.

 \end{abstract}
%

\section{Introduction}
In the context of data analysis of circular polarisation in spectral lines, 
the development of the Centre-of-Gravity technique (CoG) was initially motivated 
to cater to a method for the measurement of the magnetic field of 
spatially resolved structures present in the solar photosphere (for example sunspots), 
without recourse to detailed theoretical modeling of circular polarisation in 
line profiles \citep{semel1967}. This approach establishes a linear relation between 
the component of the magnetic field  vector projected along the line-of-sight ($B_{\rm LOS}$) 
and the relative shift between the centres of gravity of the left and right components of the 
observed circular polarisation:

\begin{equation}
\lambda_{+} -\lambda_{-} \, = \, 2 {\bar g} \Delta \lambda_{B_{LOS}},
\label{ec:COG_blos}
\end{equation}
where ${\bar g}$ is the Lander factor of the transition line, $\Delta \lambda_{B_{LOS}}$ is the 
wavelength shift due to the Zeeman splitting, and the centres of gravity for the left and right 
polarisation are respectively defined as \citep{rees1979}:
\begin{equation}
\lambda_{\pm} =  \int_{-\infty}^{\infty} \left( I_c - (I \pm V) \right) \lambda d\lambda \, \, 
 /  \, \, \int_{-\infty}^{\infty} \left( I_c - (I \pm V) \right)  d\lambda,
 \label{ec:COG_s}
\end{equation}
where $I$ and $V$ are the intensity and circular Stokes parameters, 
and $I_c$ is the (assumed unpolarised) continuum.  While from the previous 
definition the integration limits go from $-\infty$ to $\infty$, in practice 
the integration spans only around the (full) width of the line profiles; 
therefore, the selection of the integration range --which can be subjective--,  
has an important impact in the accuracy of the magnetic field measurements. 
Since Eq. (\ref{ec:COG_s}) corresponds to the first-order moment in $\lambda$, 
the CoG method  is also known as the integral method for measurements of magnetic fields
or simply as the first-order moment approach \citep[e.g.][]{mathys1989}.
Proven to be very useful, the CoG method was also applied in the stellar domain 
\citep[e.g.][]{mathys1991} to measure the mean longitudinal magnetic field
--integrated over the visible hemisphere of the star--, 
also referred as the effective magnetic field ($H_{\rm eff}$).

The CoG method was initially applied using the so-called photographic technique, 
however, with the development of new instrumentation --CCDs and spectrographs of 
high resolution with better throughputs--, 
the helpful information contained in the shape of line profiles came at this disposal. Nowadays it is 
possible to simultaneously obtain a huge number of lines in spectropolarimetric mode with 
very high resolution.

The use of mean polarised profiles resulting from the addition of multiple 
individuals lines in combination with the CoG method to infer $H_{\rm eff}$,
possible through the use of the LSD technique \citep{donati1997}, 
was a benchmark in studies
related to the stellar magnetism domain. By adding hundreds to thousands of individual lines, 
the signal-to-noise ratio of the mean circular polarised profile is increased by 
several orders of magnitude allowing the detection of extremely weak stellar magnetic fields with intensities in the order of few Gauss 
\citep[e.g.][]{marsden2014}. The use of the LSD lead to finding very 
interesting results in many types of stars and it also gave the  opportunity to shape our current 
knowledge of the stellar magnetism by observational methods using multi-line spectropolarimetric 
data analysis 
\citep[see e.g.][]{donati2009}.

For the addition of lines it is convenient to apply a variable transformation from 
wavelength to doppler velocity coordinates ($v$) \citep{semel1995}, such that the longitudinal 
stellar field would be given by \citep{mathys1989, donati1997}: 

\begin{equation}
\label{eq:heff}
H_{\rm eff} = \frac{-7.145 \times 10^{5}}{\lambda_0g_0} \frac{\int v\, \, \frac{V(v)}{I_c(v)} \, \, dv}
{\int ( 1-\frac{I(v)}{I_c(v)} \, ) \, \, dv},
\end{equation}
where $H_{\rm eff}$ is expressed in G, $v$ in ${\rm km \, s^{-1}}$. 
If only weak and unblended lines are considered,  
$\lambda_0$ (expressed in nm) and $g_0$  would correspond respectively to the means of 
the wavelengths and Lander factors of the lines employed for the establishment of 
the mean profiles.  

In fact, the CoG and the first-order moment approaches are valid under the following assumptions 
\citep{mathys1989}: 
1) an atmospheric Milne-Eddington model, 1) a weak-line formation regime (that is when the line profile
is similar in shape to the absorption coefficient ($\eta$), i.e. $\eta$ $\ll$ 1)  
and 3) are considered only weak magnetic fields (i.e. where the Zeeman splitting is much lower than 
the natural width of the line).

For the establishment of the mean profiles, if any of the 3 assumptions listed above
is not fulfilled, or if blended lines are included, 
the value of $\lambda_0 \, g_0$ has to be found by independent calibration methods. 
This important statement will be in fact the subject of this paper, namely, 
we estimate $H_{\rm eff}$ through the first-order moment 
expressed in Eq. (\ref{eq:heff}),  and we inspect different criteria used 
during the establishment of the mean profiles. Example criteria include the line
depth limit, the normalisation of the mean profiles and the integration limits.
We also investigate the role played by projected rotational velocity of the stars 
($vsini$) in the accuracy of the measurements of $H_{\rm eff}$. Finally,
all the results are discussed beyond the context of the weak field regime.

\section{Numerical tests}

The employment of the linear relation given by Eq. (\ref{eq:heff}) requires a proper 
calibration regulated solely by the product of the normalisation parameters $g_0$ and $\lambda_0$. 
In this section we will  present a series of tests in which we obtained an optimal calibration through
the use of theoretical spectra. We will denote these values by ${\mathbf{\lambda_0} \, \mathbf{g_0}}$ 
to indicate that they were found by the methodology described below.

We have used the {\sc cossam} code \citep{stift2000} to synthesise a sample of 50 polarised spectra
considering an oblique centred magnetic dipolar model \citep{stibbs1950, stift1975}.  
We have employed a solar atmospheric model: 
$T_{\rm eff} = 5750$ K, [M/H]=0, log (g) = 4.5 ${\rm cm \, s^{-2}}$, 
and microturbulence of zero, covering a wavelength range  from 365 to 1010 nm in steps of 1 ${\rm km \, s^{-1}}$.
For our first test, we adopted a slow rotator model 
in which we assigned to $vsini$ a value of 5 ${\rm km\,s^{-1}}$.
For the synthesis of each  spectrum, we have randomly varied the inclination between the 3 
principal axis of the reference system, 
namely,  the rotation axis, the magnetic dipolar axis and the line-of-sight direction. 
Considering only as free parameter these 3 angles that determine the orientation of the system,
and  setting the magnetic dipolar moment to 30 G, we obtained  that in the  synthetic sample 
the $H_{\rm eff}$ varies between  -20 and 20 G.

For the establishment of the LSD profiles we have obtained from the {\sc vald} database \citep{vald2015} 
the information required to create the mask, i.e. for each line  we retrieved from {\sc vald}
the Lander factor, the line depth ($d$) and the wavelength. 
Using a line depth limit of 0.1 with respect the continuum as
a threshold criteria, the total number of lines amounts to 8314. Of course, 
the same line list used in the mask was employed for the synthesis of the theoretical 
spectra in {\sc cossam}. Since the synthetic sample of spectra is noiseless,
we employed a cross-correlation between the mask and  each spectrum to establish the 
sample of synthetic LSD profiles, and the mask weights that we assigned for the  
Stokes I and V parameters are those included in the original LSD-paper of \cite{donati1997}:

\begin{equation}
 w_{ \textsc{i} \, i } = d_i  \, ; \\
 w_{\textsc{v} \, i } = d_i \, \lambda_i \, {\bar g_{i}} \,,
 \label{eq:ori_mask}
\end{equation}
where the index $i$ runs over the total number of lines. 

The  spectral resolution at which  the theoretical spectra was synthesised
(in wavelength steps of 1 ${\rm km\,s^{-1}}$), has to be comparable to the instrumental one.
Current observing facilities in spectropolarimetric mode have resolving powers (R)
between 55,000 ({\sc caos}) to 115,000 ({\sc harps}). We thus decided
to use  the intermediate resolution of R=65,000 that corresponds to the twin 
spectrographs {\sc espadons} and {\sc narval} (and similar to the one in {\sc boes},   R=60,000). 
In consequence, we have decreased the resolution in the synthetic spectra to constant wavelength steps 
of 1.8 ${\rm km\,s^{-1}}$ to be consistent with the adopted resolution of these two spectrographs, reducing 
the total number of lines to 8088. Finally,
to account for the instrumental broadening we convolved the spectra with a Gaussian kernel 
in which we considered a standard deviation in the Gaussian profile 
of 4.4 ${\rm km\,s^{-1}}$. 

\begin{figure}
\begin{center}
\includegraphics[width=7cm]{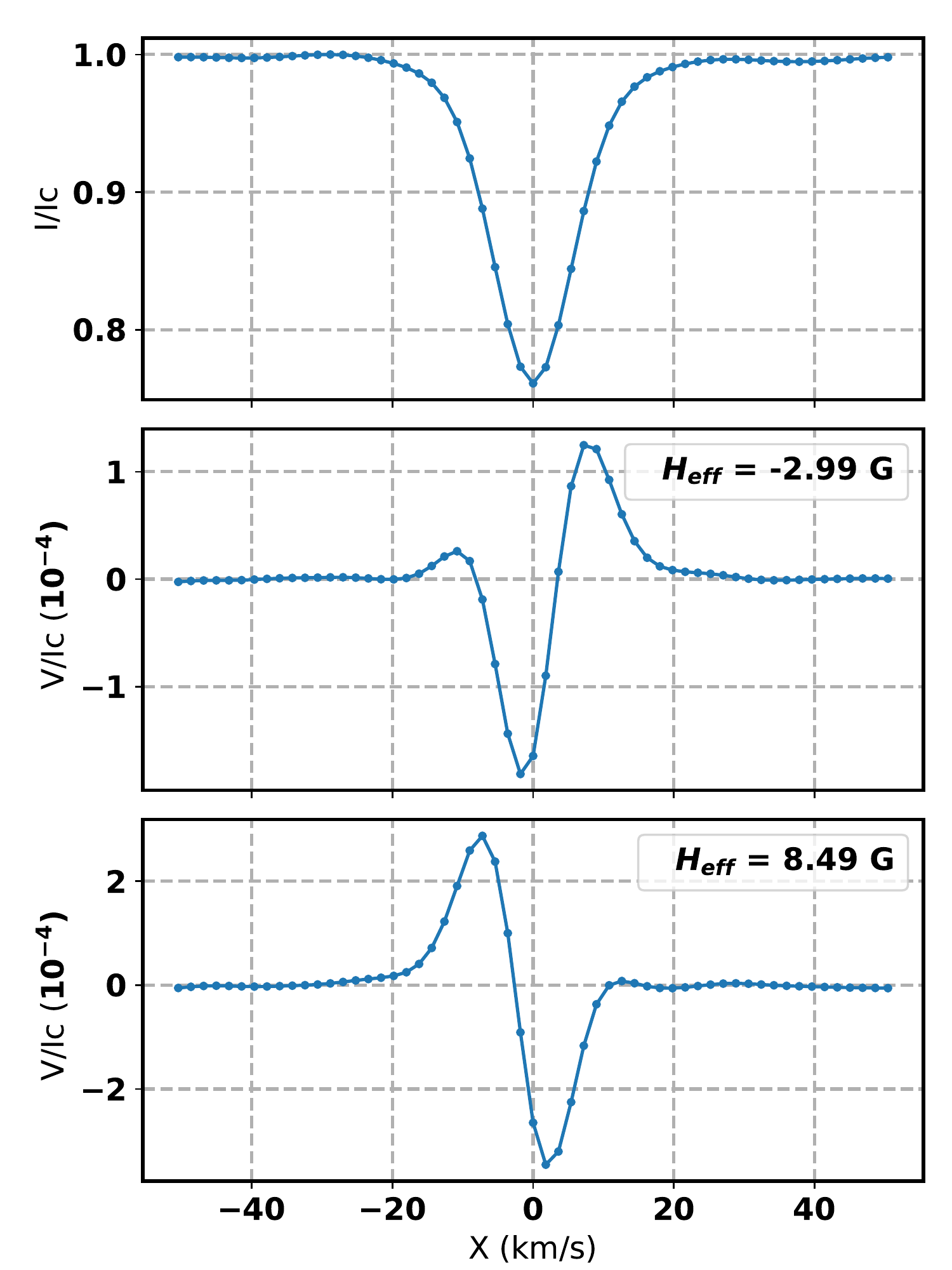}
\end{center}
\caption{Examples of LSD profiles in intensity (upper panel) and circular polarisation (middle and lower panels).
The longitudinal magnetical field for each V profile is indicated in an inner legend. 
The bin step in X --doppler coordinates-- is of 1.8 ${\rm km\,s^{-1}}$.}
\label{fig:ex_lsd}
\end{figure}

In Fig. \ref{fig:ex_lsd} we show some examples of the LSD profiles:
one for the Stokes I (upper panel) and two for Stokes V in which the respective input magnetic models
are such that the $H_{\rm eff}$ are -3.0 G (middle panel) and 8.5 G (lower panel).
Even in this ideal case where no noise was added, by visual inspection it is not clear 
if the width of the two circular polarised profiles is the same. In other words, 
regarding the shown V profiles, we must consider whether to use the same width in both cases, 
and if yes, how to find it. Before tackling this issue in detail, let us note that many  
studies have employed a visual inspection to determine the width of the observed profiles
\citep[e.g.][]{wade2000, silvester2009}, 
and also that very early on \cite{mathys1988} was remarked the importance of a proper  
determination of the line width in an analysis under a $n$-order moment approach even 
in the case of one single line.

\subsection{Profile width}

We thus proceed to inspect how the considered integration limits in the doppler space
--i.e. the LSD profiles width-- can affect the inference of $H_{\rm eff}$ when using the 
first-order moment technique. For this purpose, we have varied the width
of the  profiles from 7.2 to 97.2 ${\rm km\,s^{-1}}$
around the line centre, in steps of 3.6 ${\rm km\,s^{-1}}$.
The adopted width variation at each step corresponds to considering one more point at each 
side of the line profile, i.e., the minimal possible difference when the profile width is
varied symmetrically around the centre. For each considered profile width, we performed a
linear regression over the 50 synthetic spectra to obtain the optimal value of  
${\mathbf{\lambda_0} \, \mathbf{g_0}}$ that gives the best results to determine 
$H_{\rm eff}$ through Eq. (\ref{eq:heff}). Note that it is not possible to obtain 
separately the values ${\mathbf{\lambda_0}}$ and ${\mathbf{g_0}}$,  but only their product.

\begin{figure}
\begin{center}
\includegraphics[width=8cm]{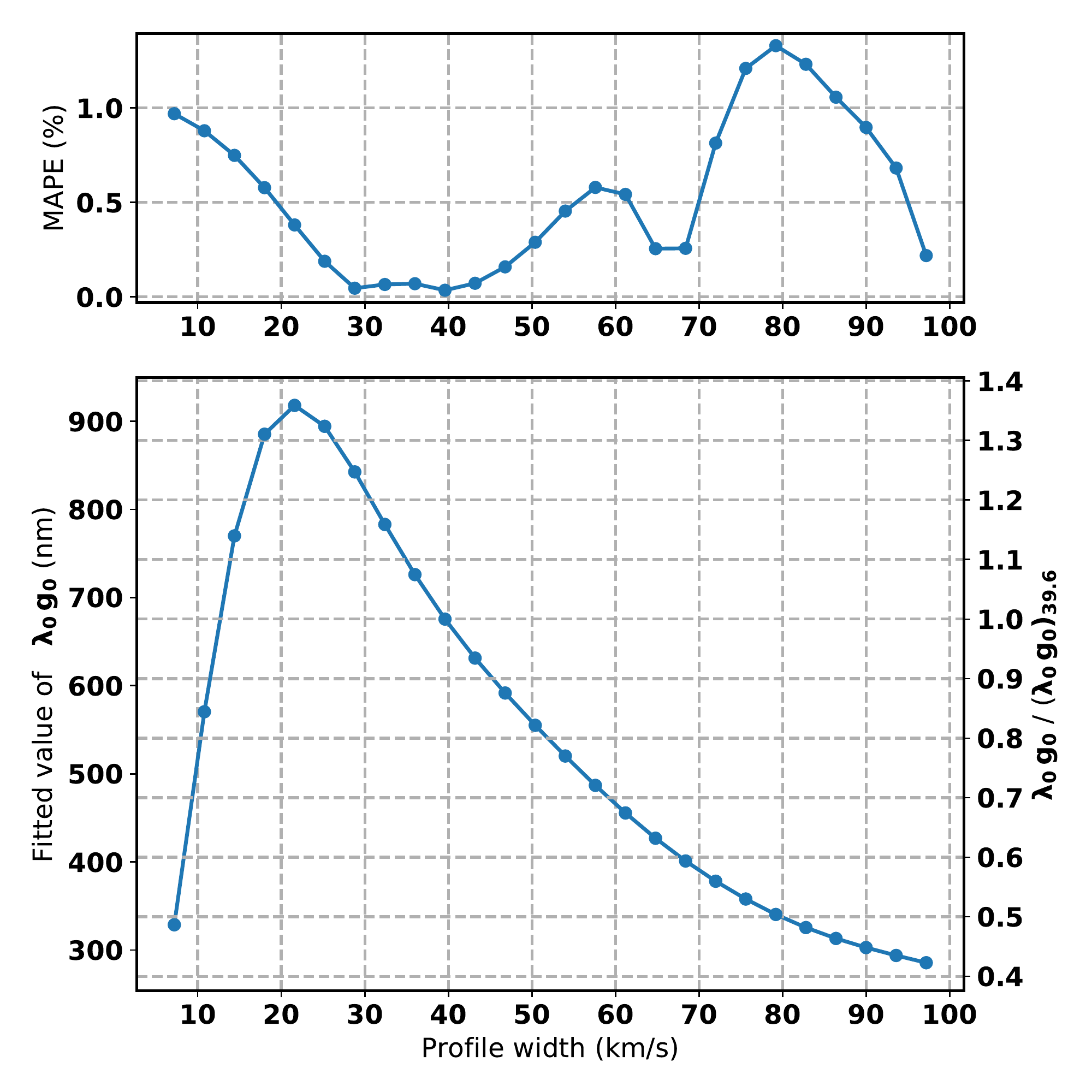}
\end{center}
\caption{Mean absolute percentual error (upper panel) and 
normalisation values (lower panel) as function of the profiles width.
In the right Y-axis of the lower panel, the values are normalised
to the case when the profile width is 39.6 ${\rm km\,s^{-1}}$.
The results were obtained analysing a sample of 50 synthetic spectra.}
\label{fig:fig1}
\end{figure}

The results are shown in Fig. \ref{fig:fig1}.
In the upper panel we show the Mean Absolute Percental Error (MAPE) obtained
by comparing the original values of  $H_{\rm eff}$ and the ones derived by the
regressions\footnote{$MAPE (\%) = \left| \frac{ H_{\rm eff}^{original}-\,H_{\rm eff}^{regression} }{H_{\rm eff}^{original}} \right|$}, 
while  the lower panel shows the respective fitted 
values of ${\mathbf{\lambda_0}} \,{\mathbf{g_0}}$,  both
as function of the integration limits (width profiles).

The first remark of this test is that for all considered profile widths, it is always possible to fit
a value of ${\mathbf{\lambda_0}} \,{\mathbf{g_0}}$ that allows us  
to infer very accurately  $H_{\rm eff}$:
the MAPE remains inferior to 1.0\% in the big majority of the cases. 
This result is quite unexpected since even underestimating
the width of the profiles up to the extreme case in which the profiles consist of only 5 
central points (from -3.6 to 3.6 ${\rm km\,s^{-1}}$), one can obtain very precise estimations 
of $H_{\rm eff}$. Analogously, 
the same behaviour is obtained when the profiles are highly overestimated, with very small MAPE values.
The fact that it is possible to consider deliberately only a part of the profiles to infer 
$H_{\rm eff}$ can be useful in some practical applications, as for example in binary systems or 
in stars surrounded by circumstellar envelopes with strong stellar winds that can generate shocks visible as bumps.
The bump produced by the  shock is in turn blended with the intensity profile of the star 
\citep[e.g.][]{sabin2015},
so to consider only a fraction of the intensity profile of the star could be of interest.

Another important result of this test is that it shows that the  value of 
$\mathbf{\lambda_0 \, g_0}$ is very sensitive to the  integration range, 
see lower panel of Fig. \ref{fig:fig1}.  The fitted values  of 
$\mathbf{\lambda_0 \, g_0}$ start at 329 nm (profile width of 7.2 ${\rm km\,s^{-1}}$) and they
increase very quickly to reach a maximum of 918 nm (profile width of 21.6 ${\rm km\,s^{-1}}$), 
to then decrease following a likely exponential-type curve, finishing at 286 nm for the broadest 
 profile width of 97.2 ${\rm km\,s^{-1}}$. 
Additionally, to illustrate the relative changes in the values of $\mathbf{\lambda_0 \, g_0}$,
we take as reference  a width of 39.6 ${\rm km\,s^{-1}}$ (from -19.8 to 19.8 ${\rm km\,s^{-1}}$), 
which seems a plausible selection
by visual inspection of the profiles in Fig. \ref{fig:ex_lsd}.
In the right Y-axis of the lower panel in Fig. \ref{fig:fig1} are shown the percentage 
variations of  $\mathbf{\lambda_0 \, g_0}$:  as example, 
if one or two more points are considered at each side of the profiles, then the 
respective errors will overestimate 
the inferred $H_{\rm eff}$ by 7.5\% and 16.0 \%. Similarly, an underestimation of  
$H_{\rm eff}$  of 6.5\% and 12.5 \% will be induced if the width of the profiles is 
reduced by one and two points respectively.

\begin{table*}
\centering
\caption{Self-consistency check of the measurements of  $H_{\rm eff}$ and estimations of errors. The mean and standard deviations 
values are denoted by <$H_{\rm eff}$> and $\sigma$, respectively. $H_{\rm eff} ^{reg}$ corresponds to the values obtained
from the regression for different profile widths.}
\label{tab:self_H}
\begin{tabular}{|l|ccccccc|r|} 
\hline
\hline
$\mathbf{\lambda_0 \, g_0} $ (nm)  &  842.6 & 782.9 & 726.0 & 675.5 & 631.3 & 591.7 & 555.0  &  \\
\textbf{Profile width} (${\rm km\,s^{-1}}$) &28.8 & 32.4 & 36.0 & 39.6 & 43.2 & 46.8 & 50.4 &  \\
\hline
\hline
& $H_{\rm eff} ^{reg}$ &$H_{\rm eff} ^{reg}$ &$H_{\rm eff} ^{reg}$ &$H_{\rm eff} ^{reg}$ &$H_{\rm eff} ^{reg}$ &$H_{\rm eff} ^{reg}$ &$H_{\rm eff} ^{reg}$ & \\
\hline
$H_{\rm eff} ^{original}$ = 2.99 G &-2.996 & -2.993 & -2.993 & -2.995 & -2.997 & -3.000 & 3.005 & <$H_{\rm eff}$> $\pm \,  3 \sigma$  = 2.99 $\pm$ 0.01 \\
\hline
$H_{\rm eff} ^{original}$ = 8.49 G& 8.492 & 8.496 & 8.497 & 8.495 & 8.491  & 8.486 & 8.479 &  <$H_{\rm eff}$> $\pm \, 3\sigma$  = 8.49 $\pm$ 0.02 \\
\hline
\end{tabular}
\end{table*}

Note that the polarised V profiles are almost zero
around $\pm$ 20 ${\rm km\,s^{-1}}$, and at first glance it could be surprising the fact that 
${\mathbf{\lambda_0}} \,{\mathbf{g_0}}$ does not remain constant when the profile width
continues to increase (integration range > |20| ${\rm km\,s^{-1}}$). 
The reason for this is due to the fact that the value of the integral in the denominator 
of Eq. (\ref{eq:heff}) continues to vary even in the regions where
the polarised signal is zero, and in consequence also 
${\mathbf{\lambda_0}} \,{\mathbf{g_0}}$ varies to get an optimal fit.

One more interesting aspect to look at is if it is possible to consider
asymmetric ranges of integration for the inference of $H_{\rm eff}$. To answer this question, 
we have resized the sample of synthetic profiles from -19.8 to 7.2 ${\rm km\,s^{-1}}$,  
and then we repeated the test. We found that in this case
the errors are considerably highers:   MAPE of  67\%. Nevertheless, we verified that the 
inversion errors decrease as the asymmetry in the profiles decreases, reaching a value of
0.5\% for the fully symmetric case (from -19.8 to 19.8 ${\rm km\,s^{-1}}$).
The conclusion is thus that the integration ranges have 
to be symmetric around the centre of profiles, but, as we showed above, 
the integration ranges do not necessarily have to include the full width of the profiles.

The results of Fig. \ref{fig:fig1} allows to infer $H_{\rm eff}$ considering
different profile widths, which in turn can be used for a check of the self-consistency 
of the measurements and to derive an associated uncertainty. Table \ref{tab:self_H},
shows the obtained values of $H_{\rm eff}$  considering 7 different profile widths  
for the two LSD profiles shown in bottom of Fig. \ref{fig:ex_lsd}.  
The top of the central columns indicate the values of  $\mathbf{\lambda_0 \, g_0}$ and the 
respective profile widths.

The extremely high precision of $H_{\rm eff}$ reported in Table \ref{tab:self_H} are due to the 
fact that the sample of LSD profiles is noiseless. However, we consider that the mean and standard 
deviation values reflect a realistic value of the measurement of $H_{\rm eff}$ and the
associated error, and could be especially useful for real observed data. 
Let us note that previous studies have shown that is very probable that the errors reported
in studies based in LSD are underestimated \citep{carroll2014, jcrv2018}.
In this sense, the {\em multi-inversions} strategy presented in Table \ref{tab:self_H}, could be a good
alternative to estimate the uncertainties. 

Note that as it is customary, the calibration 
presented for the given values in $T_{\rm eff}$ and log (g), and for the given line list,
can consider small variations; for example, in temperature 
a variation $\pm$ 125 K is still considered acceptable.

Given that the profiles used in this test are noiseless, we proceed
to consider real data but not to continue the topic of this section, 
but to discuss another two important considerations as are 
the line depth limit adopted when establishing the LSD profiles, and the inclusion of 
noise-weighted masks.

\subsection{Line depth and signals weighted by noise}

\begin{figure*}
\begin{center}
\includegraphics[width=18cm]{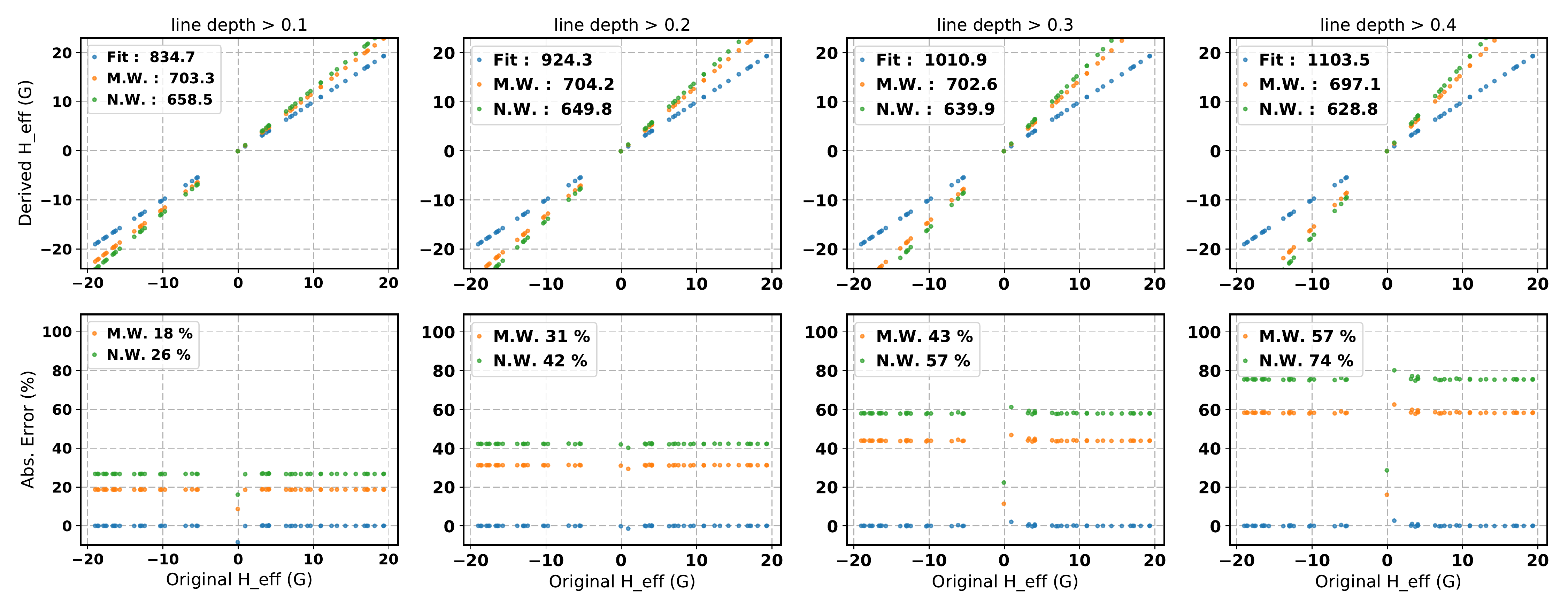}
\end{center}
\caption{Upper panels: Computation between the original and derived vales of 
$H_{\rm eff}$; in blue the optimal fit ($\mathbf{\lambda_0 \, g_0}$), 
in orange the {\em mean weights} ($\lambda_0^\textsc{mw} \, g_0^\textsc{mw}$) 
and in green the {\em noise weighted} ($\lambda_0^\textsc{nw} \, g_0^\textsc{nw}$)
normalisation. Lower panels: Absolute percentual errors; the  MAPE values 
for the {\em mean weights} and {\em noise weighted} approaches are included in the inner legends.
The line depth threshold adopted  is indicated  on top of each column. 
The number of lines considered from left to right are 7757, 5688, 4517 and 3679, respectively.}
\label{fig:ldepth}
\end{figure*}

When the analysis is applied to real data, the noise associated 
with the observations can be taken into account in different ways. 
In this section we will present some of them, which we consider the 
most employed ones.

Let us first introduce the so-called {\em mean weights} (MW) defined as \citep{marsden2014}:

\begin{equation}
MW_{ \textsc{i}}=   \frac{\sum_i S_i^2 \, \, w_{ \textsc{i}\, i}^2}{\sum_i S_i^2\, \, w_{ \textsc{i}\, i}} \, ; \\
MW_{ \textsc{v}}=   \frac{\sum_i S_i^2 \, \, w_{ \textsc{v}\, i}^2}{\sum_i S_i^2\, \, w_{ \textsc{v}\, i}} \,,
\label{eq:MW}
\end{equation}
where $S_i$ is the inverse of the incertitude derived from the data reduction process associated 
to the $ith$ line (i.e. the signal-to-noise ratio of the $ith$ line), 
and the weights $w_{\textsc{i} \, i}$ and $w_{\textsc{v} \, i}$ are given by :

\begin{equation}
 w_{ \textsc{i} \, i } = \frac{d_i}{d_0^n}  \, ; \\
 w_{\textsc{v} \, i } = \frac{d_i \, \lambda_i \, {\bar g_{i}} } {d_0^n\, \lambda_0^n\,  {g_{0}^n}} \, ,
 \label{eq:mask_w}
\end{equation}
where  the parameters $d_0^n$, $\lambda_0^n$ and $g_0^n$ are referred as 
the {\em normalisation values} \citep[e.g.][]{kochu2010} or {\em scaling factors} \citep[e.g.][]{petit2014}.
We have here adopted a different notation of the {\em normalisation values}, normally expressed 
also as $\lambda_0$ and ${g}_0$. The reason of our notation is to avoid confusion because  the 
{\em normalisation values}  $\lambda_0^n$ and $g_0^n$   
of Eq. (\ref{eq:mask_w}) are not always the same as those 
used to derive $H_{\rm eff}$ in Eq. (\ref{eq:heff}), i.e.,
$\lambda_0^n g_0^n  \ne \lambda_0 g_0$.

Nowadays,  when the line mask of the LSD profiles are established it is normally chosen that 
the {\em mean weights} are numerically equal or very close to unity. It is then  
assumed that the amplitude of the resulting  LSD profiles is properly  
normalised by the definition adopted 
for the {\em mean weights}, and in consequence the normalisation parameters are directly 
used to measure $H_{\rm eff}$ in Eq. (\ref{eq:heff}),  i.e. $\lambda_0^n g_0^n  = \lambda_0 g_0$.

As we mentioned, the {\em mean weights} are not the only way to normalise the LSD profiles.
In fact, in the original LSD-paper of \cite{donati1997}, the authors proposed
to scale the amplitudes of the profiles to $\lambda_0^n g_0^n d_0^n$ = 500 nm and, 
to use the mean values of $\lambda_i$ and $\bar{g_i}$ to measure $H_{\rm eff}$ in Eq. (\ref{eq:heff}),
i.e., $\lambda_0$ = <$\lambda_i$>  and $g_0$ = <$\bar{g_i}$>.   This way of normalising the LSD profiles
was used for some years, but the normalisation values changed  by introducing a new 
constraint:  $d_0^n =0.7$  and at the same time $\lambda_0^n g_0^n d_0^n$ = 500 nm
\citep[e.g.][and others]{wade2000,shorlin2002, donati2003}. It is important
to mention that this approach to normalising the LSD profiles is not used anymore.

Alternatively, a {\em noise weighted} (NW) definition of the 
average values of  $\lambda_0$ and $g_0$ has also been used in which the signal-to-noise ratio of the 
lines is included by the following expressions \citep[e.g][]{kochu2010, grunhut2013}:

\begin{equation}
\lambda_0 = \frac{\sum_i S_i^2 \, \, \lambda_{i}} {\sum_i S_i^2} \, ; \\
\bar{g}_0 = \frac{\sum_i S_i^2 \, \, \bar{g}_{i}} {\sum_i S_i^2}\, .
\label{eq:NW}
\end{equation}
In this {\em noise weighted} approach  the average values are used to both, 
normalise the amplitude of the LSD profiles and to measure $H_{\rm eff}$,
i.e., $\lambda_0$ = $\lambda_0^n$ and $g_0$ = $g_0^n$. 

We next compare two of the  normalisation strategies presented above, namely, 
the {\em mean weighted} and the {\em noise weighted}. For the latter,
the normalisation values, denoted by $\lambda_0^\textsc{nw} \, g_0^\textsc{nw}$,
will be directly obtained through Eq. (\ref{eq:NW}). For the former, given that 
$MW_{\textsc{i}}$ and $MW_{\textsc{v}}$ are by construction equal to 1, 
the normalisation values, denoted by $\lambda_0^\textsc{mw} \, g_0^\textsc{mw}$, 
are given by:

\begin{equation}
d_0^\textsc{mw} = \frac{\sum_i S_i^2 \, \, d_{i}^2} {\sum_i S_i^2 d_i} \, ; \\
\lambda_0^\textsc{mw} g_0^\textsc{nw} = \frac{\sum_i S_i^2 \, \, d_i^2 \lambda_i^2 \bar{g}_{i}^2} {d_0 \sum_i S_i^2 d_i \lambda_i \bar{g}_{i}}\, .
\label{eq:l0g0NW}
\end{equation}

Additionally, we will include different values of the line depth limit, which is  another 
important  criteria when establishing the LSD profiles. In the published studies based in LSD, 
different line depth threshold values  have been employed, from 5\% to 40 \% with respect 
to the continuum,  but the most commonly used ones are 10\%, 20\% and 40\%.  We now proceed to quantify how much the value of the 
product of $\lambda_0 \, g_0$ varies when considering different line depth limits.

For this purpose, it is necessary to consider noise-affected data. Despite that it is always possible
to model the noise following random or Poisson distributions, here we prefer to use real
data. We have therefore obtained from {\em PolarBase} \citep{petit2014} the files associated 
with the solar twin type 
star HD63433, observed with the ESPaDOnS spectrograph at the CFHT telescope the 10 January 2010
\footnote{The block reference number of this observation in the PolarBase database is 8450.}.
In the files associated with the data reduction, we find the uncertainty in intensity
that are the inverse of the $S_i$ values required in Eqs. (\ref{eq:NW}) and (\ref{eq:l0g0NW}).
We applied a linear interpolation to the wavelength sampling of the observed data 
to match the exact  $w_i$ values of the mask, which is a standard procedure.
By considering the same spectral region in the synthetic sample and the
observed data, the number of lines reduces to 7757 (with a line depth > 0.1).
Besides, we have considered a value of $vsini$ = 7.0 ${\rm km\,s^{-1}}$ in the 
synthetic sample of spectra to be consistent with the value reported by \cite{valenti2005}
for the projected rotational velocity of HD63433.

Finally, in the previous section we have shown that $\lambda_0 \,g_0 $
is dependent of the  integration range in the doppler space (profile width).
Therefore for this numerical test, we have fixed the profile width to 28.8 ${\rm km\,s^{-1}}$,
as indicated in the reduction log files of this observation. The integration 
limits that we considered vary from  -14.4 to 14.4  ${\rm km\,s^{-1}}$  in the rest frame of the star. 
The results are shown in Fig. \ref{fig:ldepth}.

The first  remark of the results of this figure is that both, the {\em mean weights} and the {\em noise weighted}
normalisation values, remain almost constant. By changing the threshold  of the 
line depth from 0.1 to 0.4, the values of $\lambda_0^\textsc{mw} \, g_0^\textsc{mw}$
shows  a decrease of only  $\sim$ 1\% (from 703 to 697 nm), 
while for $\lambda_0^\textsc{mw} \, g_0^\textsc{mw}$ the decrement
is $\sim$ 5\% (from 659 to 629 nm). 
On the contrary, the  optimal fit obtained for $\mathbf{\lambda_0 \, g_0}$ increases in 32\%  (from 835 to 1104 nm).
In consequence, it is shown from left to right in the lower panels of Fig. \ref{fig:ldepth}, 
that the errors increase from  18\% to 56\% for the {\em mean weight} estimations of $H_{\rm eff}$,
while for the {\em noise weight} approach the errors are even higher passing from  26\% to 74\%.

The conclusion of this test is that for each observation, or equivalently for each set of
given values of  $S_i$, there is only one value of line depth that fulfils
$\lambda_0 \, g_0$ = $\lambda_0^{\textsc{MW}} \, g_0^{\textsc{MW}}$, and analogously
only one other  such that $\lambda_0 \, g_0$ = $\lambda_0^{\textsc{NW}} \, g_0^{\textsc{NW}}$.
For the case presented here, those two values are inferior to a line depth limit of 0.1.
Therefore, it is not convenient to search for which line depth value 
the  {\em mean weights} or {\em noise weighted} are good normalisations, but the contrary, that given 
any limit for the line depth, the values of  $\lambda_0 \, g_0$ must be found through 
synthetic spectra, as we have proceeded here. Currently, some studies indicate that $\lambda_0 \, g_0$ 
are derived through synthetic spectra, however the procedure is not described and 
the values of $\lambda_0 \, g_0$ are in general only announced. 


\subsection{Vsini}
In this section, we will focus on inspecting whether the estimations of the stellar longitudinal
magnetic field can be affected by the rotation of the star, an effect normally ignored,
and if yes how important it is to consider the $vsini$ value in the synthetic sample of  
spectra. In fact, by increasing the projected rotational speed of the star,
two important effects appear. First, the blending of the lines increases, 
and second, the weak-line regime is less justified up to the case in which 
the shape of the line profile is rotationally dominated. With the aim of investigating 
the resultant interplay of these two effects, we present an estimation of the $\lambda_0 \, g_0$ values as 
function of $vsini$.

We have considered the same solar atmospheric model as in the previous tests,
with the only difference that now the projected rotational values are varied
from 0 to 50 ${\rm km\, s^{-1}}$ in steps of 5 ${\rm km\, s^{-1}}$.
For each of these $vsini$ values, we synthesised a sample of 50 spectra to then 
fit a linear regression  obtaining in each case  $\mathbf{\lambda_0 \, g_0}$.

\begin{figure}
\begin{center}
\includegraphics[width=7cm]{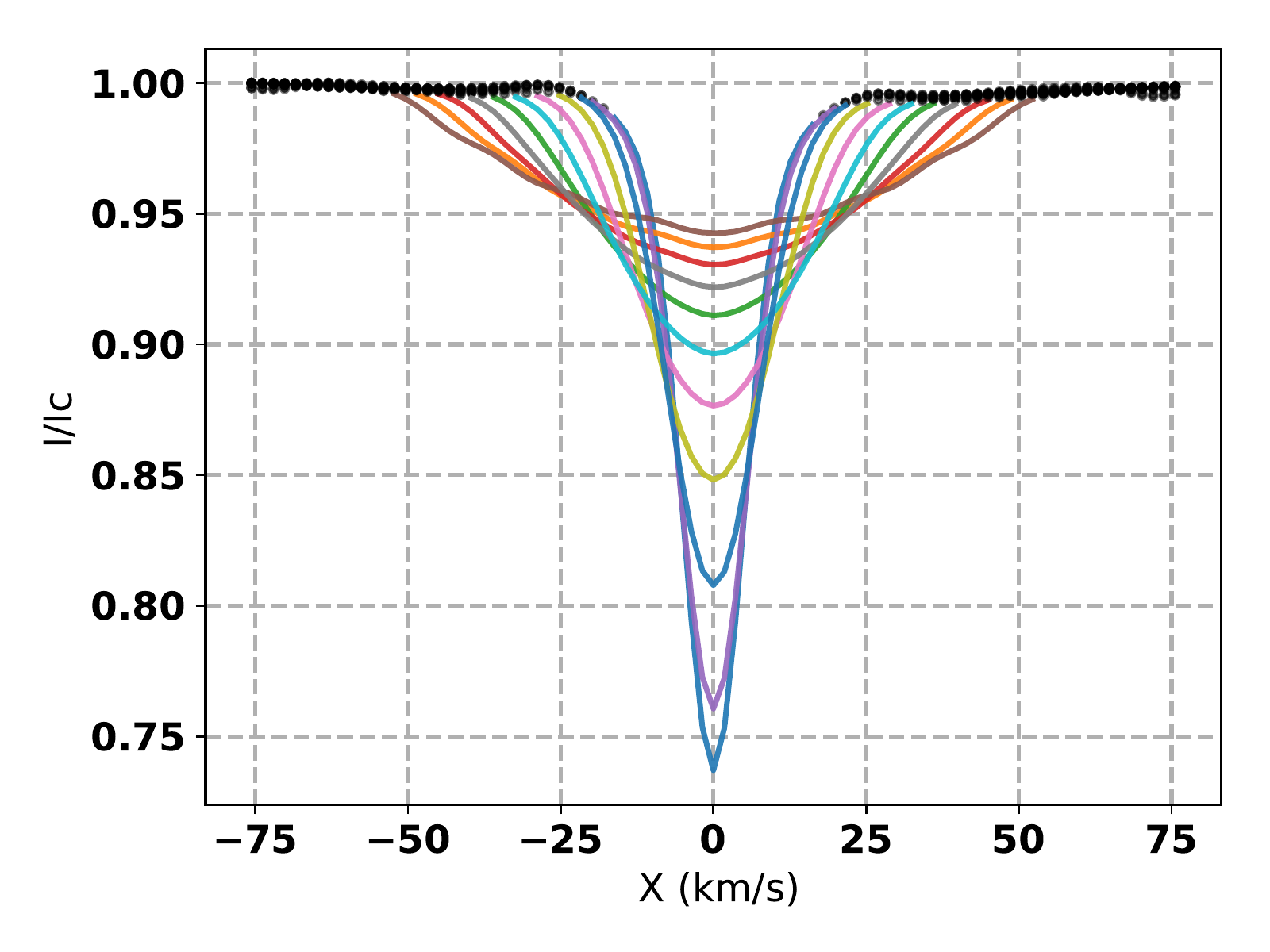}
\end{center}
\caption{In colour the profile widths that determine the integration limits 
to infer $H_{\rm eff}$. The profiles correspond to different values of $vsini$ 
that vary from 0 to 50 ${\rm km\, s^{-1}}$ in constant steps of 5 ${\rm km\, s^{-1}}$.}
\label{fig:lsd_vsini}
\end{figure}

Before continuing, it is essential to define the integration limits
for each of the $vsini$ values. We have visually inspected simultaneously both
Stokes profiles, I and V, to define their widths. Strategies other than visual
inspection could be adopted but for the main purpose of this test (to determine
the variation of $\mathbf{\lambda_0 \, g_0}$ as function of $vsini$), we consider that 
the employed strategy is enough. Fig. \ref{fig:lsd_vsini} shows in colour the 
selected profile widths, 
and in black the remaining part of the profiles for all the rotational values.

\begin{figure}
\begin{center}
\includegraphics[width=8cm]{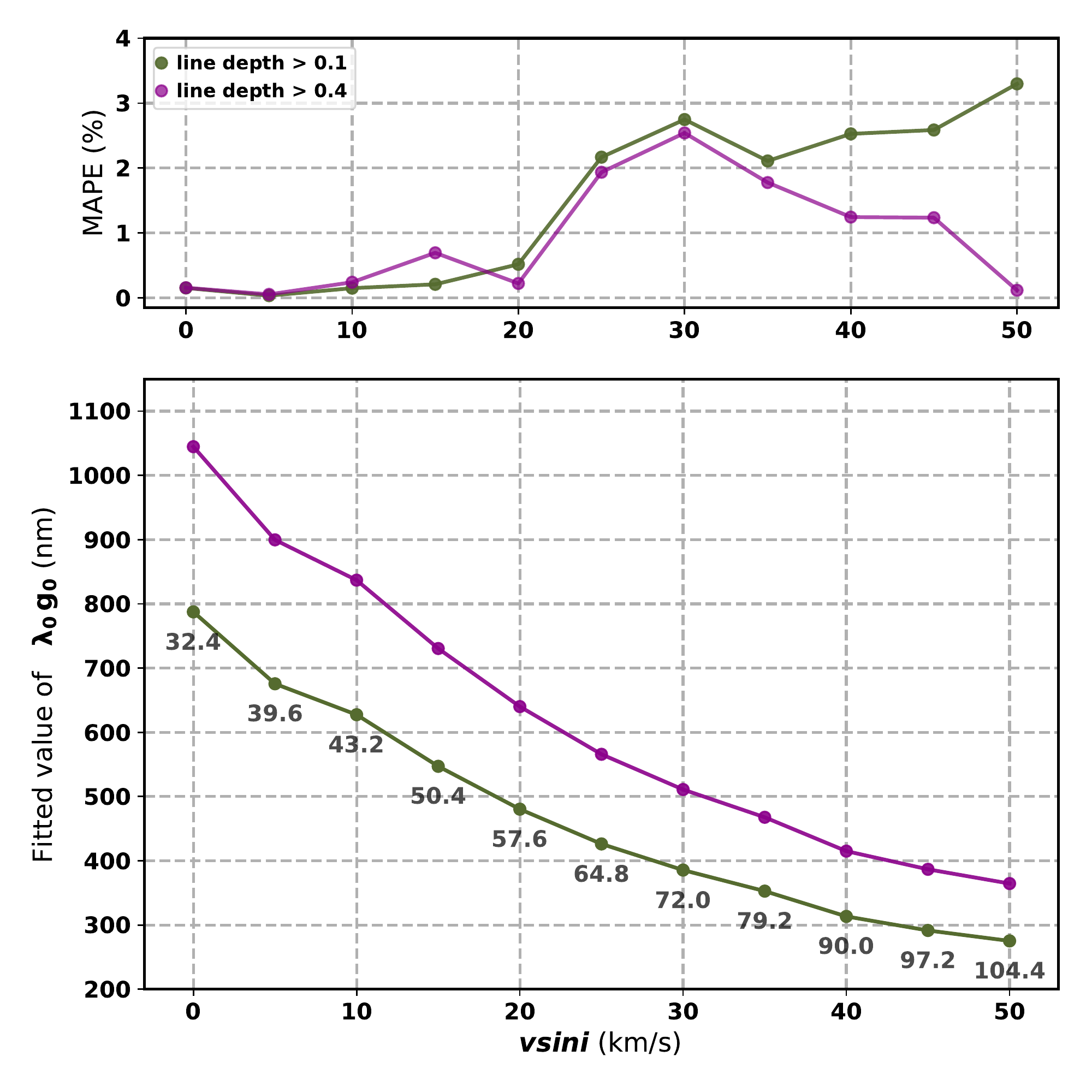}
\end{center}
\caption{Mean absolute percentual error (upper panel) and 
normalisation values (lower panel) as function of 
$vsini$, considering line depth limits of 0.1 (in green) and 0.4 (in magenta). The numbers below
the green line in the lower panel indicate the considered width of the profiles.}
\label{fig:cog_vsini}
\end{figure}

For comparison purposes, in the establishment of the LSD profiles we have considered 
two different values of the line depth limit,  0.1  and 0.4. 
The results shown in Fig. \ref{fig:cog_vsini} are presented as before:  
the upper panel shows the precision of the inversions quantified through the MAPE
values and the lower one shows the fitted values $\mathbf{\lambda_0 \, g_0}$, both 
as function of $vsini$.  For completeness, below the curve of line depth limit of 
0.1 of the lower panel, we have included the profile width employed at each rotational
value: For example, for a $vsini$ of 10 ${\rm km\, s^{-1}}$, the
indicated profile width is 43.2 ${\rm km\, s^{-1}}$, which means that the integration range goes from 
-21.6 to 21.6 ${\rm km\, s^{-1}}$ in the rest frame of the star.

\begin{figure*}
\begin{center}
\includegraphics[width=18cm]{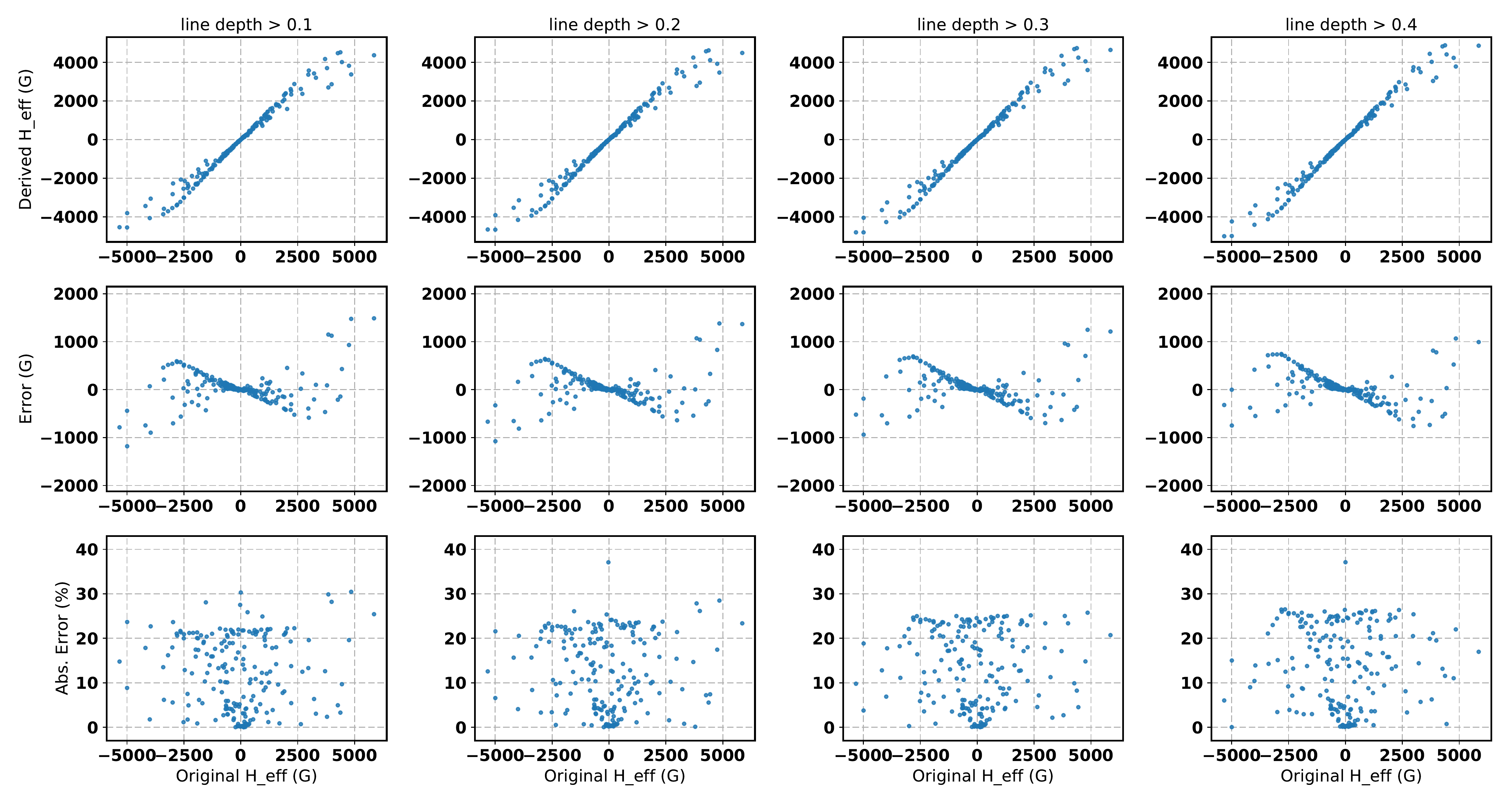}
\end{center}
\caption{Comparison of the input and inferred $H_{\rm eff}$ values (upper panels). The 
errors are shown in units of Gauss (middle panels) and in absolute percentage (lower panels).}

\label{fig:Cog_heff}
\end{figure*}

The results from Fig. \ref{fig:cog_vsini} indicate that for all the rotational values 
the retrieved values of $H_{\rm eff}$ are very good:
except in one case the MAPE is always less than 3\%. Therefore, with this test it is shown that
even in the case with the strongest rotation velocity  of 50 ${\rm km\, s^{-1}}$, whose
shape profile is clearly dominated by the rotation and the blending process are the highers,
the first-order moment is still a very good tool to measure the longitudinal magnetic
field.  However,  good precision of estimations of $H_{\rm eff}$ requires an appropriate
value  $\mathbf{\lambda_0 \, g_0}$ for each rotational value. In fact, the normalisation value 
changes  by a factor close to 3 when 
passing from extreme to the other of the $vsini$ values, from  0 to 50 ${\rm km\, s^{-1}}$. 
In other words, 
if the $\mathbf{\lambda_0 \, g_0}$ corresponding to a null rotational value is employed
to infer $H_{\rm eff}$ for a moderate fast rotator star with a $vsini$ of 50 ${\rm km\, s^{-1}}$, 
our results indicate the magnetic fields will be underestimated by 287\% 
(the same percentage was found for both line depth limits of 0.1 and 0.4). 
The conclusion of this test  is therefore that it is mandatory to consider the projected rotational velocity of the stars 
when determining the normalisation values $\lambda_0 \, g_0$, something that has been neglected up to today.

Please note that similar results have been obtained not on the basis of the
integral form (Eq. \ref{eq:heff}) but rather on the derivate one. 
Recently, using the slope method, \cite{scalia2017} have also shown that the 
the errors of the longitudinal magnetic field increase when the $vsini$ increases,
while \cite{leone2017} indicate that $H_{\rm eff}$ is properly estimated only 
for rotational velocities lower than  12 ${\rm km\, s^{-1}}$.

\subsection{Magnetic field intensities}
 
In the previous sections we have purposely performed the tests in a regime where
the weak magnetic field assumption was assured : $ | H_{\rm eff} |$ $<$ 20 G, and
the magnetic moment was fixed to 30 G. We are now interested in considering stronger 
intensities in the magnetic fields.

For the next numerical exercise, we synthesised a sample of 200 stellar spectra with the same
atmospheric model as before, adopting a $vsini$ value of 5 ${\rm km\, s^{-1}}$,
but now the magnetic moment was increased up to two more orders of magnitude, 
varying between 0.1 and 10 kG. The reason that we have considered a higher number of 
synthetic spectra is to better sample $H_{\rm eff}$, which varies in a wider range of intensities,
between -6 and 6 kG.

For the inference of $H_{\rm eff}$, in Eq. (\ref{eq:heff}) we have used the value 
 $\mathbf{\lambda_0 \, g_0}$ = 675.5 nm  that was found 
previously when  only weak magnetic fields were considered, the line depth limit was 0.1,
 $vsini$ =  5 ${\rm km\, s^{-1}}$, and 
the integration limits varied between -19.6 to 19.6 ${\rm km\, s^{-1}}$, see Table \ref{tab:self_H}.
The sample of spectra for this test was in consequence inverted using this same
range of the integration. For completeness, we have also considered the line depth threshold values
of 0.2, 0.3 and 0.4. For each of these cases, we have considered the values 
of $\mathbf{\lambda_0 \, g_0}$ derived previously under a weak magnetic field regime, 
which are 749.0, 822.6 and 899.6 nm, respectively. The results are shown in Fig. \ref{fig:Cog_heff}.
Before to discuss the results, it is important to note that these do not depend
on the integration range. We have verified that by changing the profiles width (integration range),
the results are in essence the same.

%
%
%

The first remark of this test is that the results are the same despite the
line depth cut-off value considered in the mask (indicated in top each column). 
The second is that it is not possible (in function of $H_{\rm eff}$) 
to constrain where the weak magnetic field  regime breaks down, given that for 
both the  weakest  (< 500 G) and the strongest 
 (> 2 kG)  intensities, it is possible achieve very accurate values of 
 $H_{\rm eff}$ in some cases, but the errors can reach 25\% to 30\% in others.
Thus, the conclusion of this test is that with a dipolar magnetic configuration,  
and considering random orientations of the principal axis of the system,
it is unfortunately not possible to a priori set a limit to the validity of the weak field
approximation using as constrain $H_{\rm eff}$.

\begin{figure}
\begin{center}
\includegraphics[width=7.5cm]{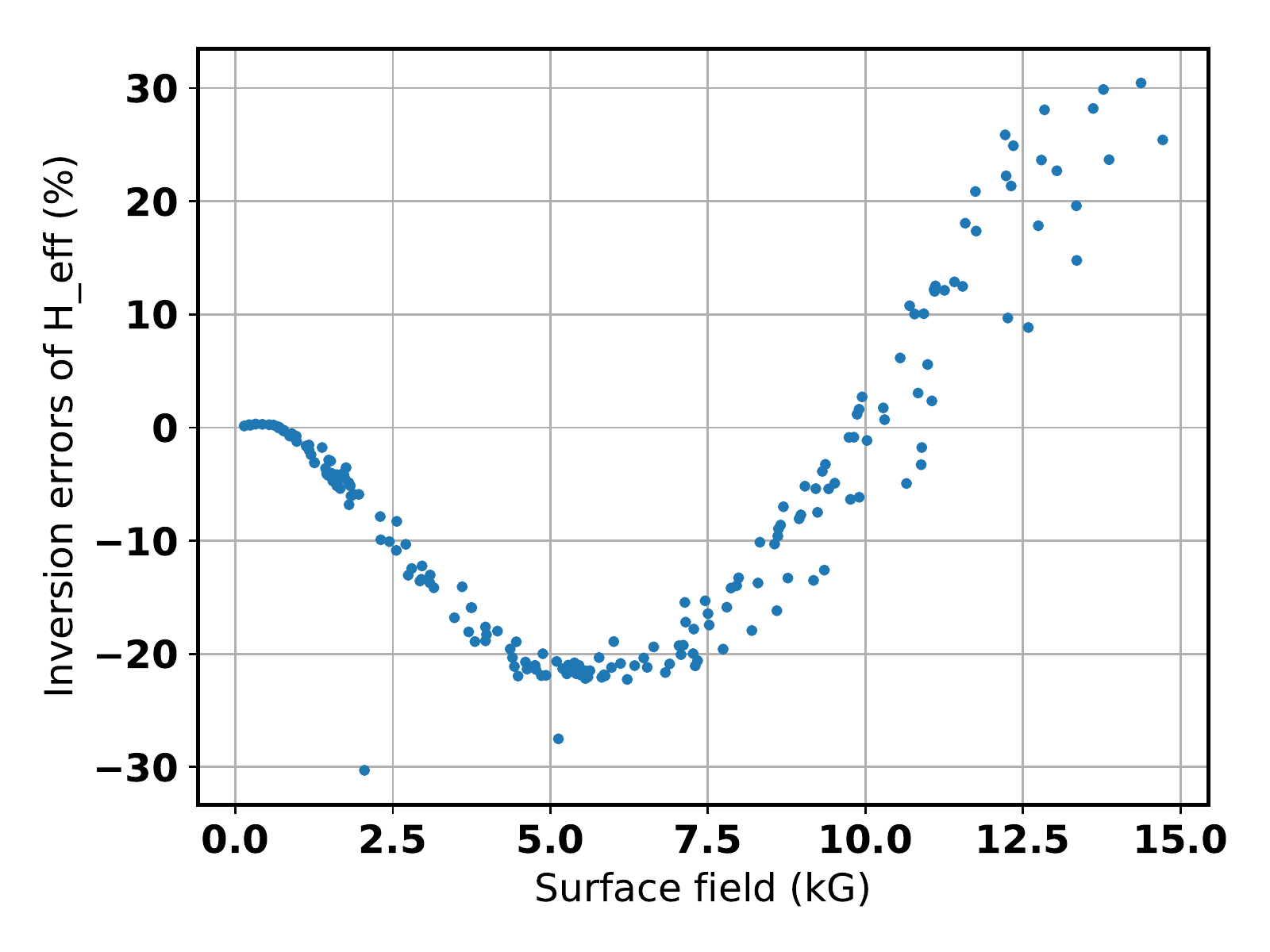}
\end{center}
\caption{Percentual errors of $H_{\rm eff}$, given by $\frac{H_{\rm eff}^{ori} - H_{\rm eff}^{reg}}{H_{\rm eff}^{ori}}$, 
as function of magnetic surface field. The results correspond to a line depth limit of 0.1.}
\label{fig:hsup_vs_heff}
\end{figure}

A clear interpretation of the results of Fig. \ref{fig:Cog_heff} can be obtained in terms
of the surface magnetic field  ($H_{\rm surf}$), defined as the mean 
of the local magnetic field moduli. In Fig. \ref{fig:hsup_vs_heff}, we show the inversion errors
of $H_{\rm eff}$ as function of $H_{\rm surf}$. In this case, it is evident that when the
weak magnetic field  regime is assured  the inversions errors are extremely low. However, 
around 1 kG the weak magnetic field assumption  starts to not be valid
and in consequence the inversion errors --which overestimate $H_{\rm eff}$-- 
begin to increase, reaching a maximum of 
20\%  for surface fields around 5 kG. Surprisingly, when the
surface intensities continue to increase, the overestimation of
$H_{\rm eff}$ starts to decrease, is close to zero around 10 kG and then
inversion errors begin to underestimate $H_{\rm eff}$. There is no clear explanation
for this empirical behaviour, and more tests should be carried out
to inspect it in detail, something that is beyond the scope
of this study.

Unfortunately once more, the fact that inversions error of the magnetic 
longitudinal field can be constrained by the surface field intensity, it is not useful
in practice for the analysis of snapshot spectropolarimetric observations, since for this 
it is required to know the distribution of the local magnetic
fields over the star. However, the results of Fig. \ref{fig:Cog_heff} are
helpful for the magnetic imaging technique.


In order to disentangle the limitations of LSD profiles separately of those of the 
the first-order moment approach, we will now employ other technique to derive
$H_{\rm eff}$, namely, machine learning algorithms.
In \cite{jcrv2018}, we have shown that these algorithms are indeed very accurate 
for measurements of longitudinal magnetic fields, with MAPE values similar to the ones
found in previous sections.

\begin{figure}
\begin{center}
\includegraphics[width=9cm]{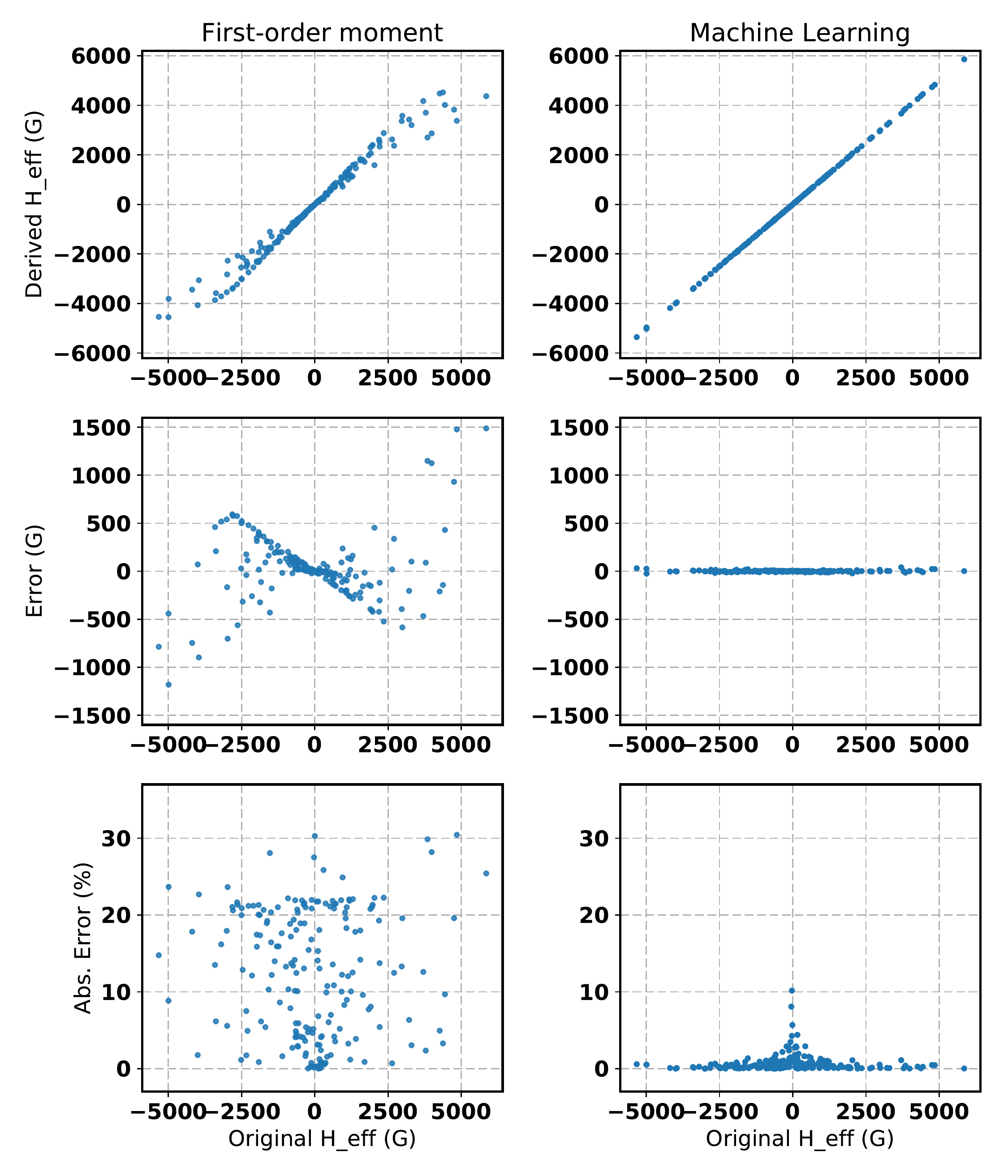}
\end{center}
\caption{Comparison of the inversions results of both techniques, 
the first-order moment approach (left columns) and the Artificial Neuronal Network 
(right columns). The sample of synthetic profiles
is the same in both columns, and corresponds to the case of a line depth limit of 0.1.}
\label{fig:cog_vs_ml}
\end{figure}

Using the same sample of 200 synthetic spectra of Figs. \ref{fig:Cog_heff} and 
\ref{fig:hsup_vs_heff}, we have trained an Artificial Neuronal Network (ANN).
The proper functioning of the ANN was performed
using a K-fold = 5 validation test, which means that the full sample of 
200 spectra is divided in five subsamples. Then, each subsample (consisting
of 40 spectra) is used to train the ANN, and the inversions are performed
over the remaining 160 spectra. This, process is repeated for all subsamples,
and the obtained validation coefficient was 0.99 \citep[][in particular,
all the technical details about the ANN are given in the appendix of that work]{jcrv2018}. 
The results obtained with the ANN trained with the 
LSD profiles are shown in Fig. \ref{fig:cog_vs_ml}.

It is clear from the right columns of Fig. \ref{fig:cog_vs_ml} that to use a
sample of LSD profiles to train a machine learning algorithm allows to determine $H_{\rm eff}$
very precisely for all magnetic intensities, from -6 to 6 kG. In consequence,
it is demonstrated with this test that the inaccuracy of the results
of Figs. \ref{fig:Cog_heff} and \ref{fig:hsup_vs_heff} is due to the use of the
first-order moment approach and not to the LSD profiles.

\section{Conclusions}

The present study is not the first one to highlight the crucial role played by 
normalisation parameters when measuring the magnetic stellar field
through the LSD profiles. \cite{kochu2010} have already stated that the same 
values to normalise the LSD profiles must be the same ones  used in the first-order
moment formula, i.e. $\lambda_0 \, g_0$ = $\lambda_0^n g_0^n $. That is in fact 
what we found in the controlled tests of the previous section.
We stress that it is always possible to use any of the mean strategies -- 
either the simple mean $<\lambda_i><\bar{g_i}>$, the {\em mean weighted} 
or the mean {\em noise weighted}--, but for these approaches it is required 
to a priori find the value of line depth for which the mean values will be equal
 to $\lambda_0 \, g_0$. 

Regardless of the employed normalisation strategy and considering 
that $\lambda_0 \, g_0$ are found through synthetic spectra, there are still
two important considerations normally set aside. The first one, already discussed, is that 
the normalisation values are dependent on $vsini$. One of the reasons that the dependence 
in $vsini$ was ignored is due to the fact that it was assumed that all lines are autosimilars, 
something that is clearly not the case for different rotational values with different line blendings.
The second one is that many recent publications have analysed samples of stars in which the mask of each 
star is carefully established but  the normalisation values  are the same for all the stars, 
despite that in the sample the stars have different atmospheric parameters as $T_{\rm eff}$, log (g),
$vsini$, etc, \citep[e.g.][]{alecian2013, villebrun2019, hill2019}.

Our results concerning the integration range of Eq. (\ref{eq:heff}), is contrary to what other
previous studies announced where it was stated that there is only one 
{\em appropriate} width of the LSD profiles that allows to 
correctly determine $H_{\rm eff}$ \citep[e.g.][]{neiner2012}. Here, we have shown 
that even taking  a small portion of the full width of the profiles, or to highly 
overestimate the width of the profiles, is useful to  correctly determine $H_{\rm eff}$ (if 
$\lambda_0 \, g_0$ is previously found for each profile width
and if the integration limits are symmetrical around the centre of profiles).
Moreover,  given that it is possible to use different profile
widths, i.e. different integration limits  in Eq. (\ref{eq:heff}), 
we proposed a method to estimate the incertitude of the measurements 
using the {\em multi-inversions} strategy (see Table \ref{tab:self_H}).

In fact, the main conclusion of this work is that the use of the first-order moment technique 
for the measurement of $H_{\rm eff}$ from multi-line LSD profiles is a very robust approach 
if and only if the  parameters $\lambda_0 \, g_0$ of Eq. (\ref{eq:heff}) are properly determined
and provided that the weak magnetic field regime is fulfilled. 
We showed that a sound methodology to find  $\lambda_0 \, g_0$ is through the use of a small sample 
of theoretical spectra calculated with the  physical parameters as close 
as possible to the data 
one wish to analyse. In this respect, please note that we have inspected only those physical 
parameters that we consider to be  the ones which have more impact in the first-order approach, 
but we have left out many other parameters such as micro and macro turbulence, 
metallicity, log (g), and others, which  we considered to have a less impact on 
the linear relation given by Eq. (\ref{eq:heff}).

There is no doubt that with the results shown in this work some of the previous reported 
measurements of stellar magnetic fields, through a combined analysis of LSD profiles and the 
first-order moment method, deserve to be revised. More importantly, new studies
using the first-order moment technique must properly calibrate $\lambda_0 \, g_0$ in order to give
more accuracy to the results. This fact is likely crucial for fast rotator stars,
in which it seems that the magnetic fields reported for these stars have been systematically
underestimated. With case study presented here where a solar atmospheric model was considered, 
the underestimation reached almost  300\% around 
$vsini\, \sim \,  50 \, {\rm km \,  s^{-1}}$. The final conclusion 
is  that in general the intensities  of magnetic fields in fast rotators stars, 
where $H_{\rm eff}$ has been measured through the first-order moment and the LSD profiles,  
is expected to be more intense than believed.

Finally, we also showed that very good measurements of $H_{\rm eff}$ are no
longer possible if the magnetic weak field assumption is not valid. With 
the magnetic dipolar model employed to establish the polarised synthetic samples,
we could not find a critical value of $H_{\rm eff}$ from which the weak field approximation 
breaks down. In other words, extremely weak measured values  of $H_{\rm eff}$ do not
assure the weak field regime. This fact is a consequence of the well known effect of attenuation
of circular polarised signals due to the  balance of positive and negative polarities of the 
magnetic field over the visible hemisphere of the star.

We also showed how to overcome this problem by using alternative methods
as are the machine learning algorithms \citep{jcrv2018}. Using the same sample of LSD 
profiles, we could properly infer the values of $H_{\rm eff}$ 
for the full sample of LSD profiles,  including strong intensities in the order of
kG (Fig. \ref{fig:cog_vs_ml}).
This demonstrates that the main constrain when deriving the stellar longitudinal magnetic
fields are not the LSD profiles, but the use of the first-order moment approach, which
is based in assumptions that can be very restrictives in practice given that the value
of $H_{\rm eff}$ does not allow a piori determine if the weak field approximation 
is assured.

\section*{Acknowledgements}
The author thanks to Franco Leone for helpful discussions that helped to improve
the content of the manuscript. This study has been supported by UNAM through the PAPIIT 
grant number IN103320.

\bibliographystyle{mnras}
\bibliography{ref_cog}

\begin{thebibliography}{}
\makeatletter
\relax
\def\mn@urlcharsother{\let\do\@makeother \do\$\do\&\do\#\do\^\do\_\do\%\do\~}
\def\mn@doi{\begingroup\mn@urlcharsother \@ifnextchar [ {\mn@doi@}
  {\mn@doi@[]}}
\def\mn@doi@[#1]#2{\def\@tempa{#1}\ifx\@tempa\@empty \href
  {http://dx.doi.org/#2} {doi:#2}\else \href {http://dx.doi.org/#2} {#1}\fi
  \endgroup}
\def\mn@eprint#1#2{\mn@eprint@#1:#2::\@nil}
\def\mn@eprint@arXiv#1{\href {http://arxiv.org/abs/#1} {{\tt arXiv:#1}}}
\def\mn@eprint@dblp#1{\href {http://dblp.uni-trier.de/rec/bibtex/#1.xml}
  {dblp:#1}}
\def\mn@eprint@#1:#2:#3:#4\@nil{\def\@tempa {#1}\def\@tempb {#2}\def\@tempc
  {#3}\ifx \@tempc \@empty \let \@tempc \@tempb \let \@tempb \@tempa \fi \ifx
  \@tempb \@empty \def\@tempb {arXiv}\fi \@ifundefined
  {mn@eprint@\@tempb}{\@tempb:\@tempc}{\expandafter \expandafter \csname
  mn@eprint@\@tempb\endcsname \expandafter{\@tempc}}}

\bibitem[\protect\citeauthoryear{{Alecian} et~al.,}{{Alecian}
  et~al.}{2013}]{alecian2013}
{Alecian} E.,  et~al., 2013, \mn@doi [\mnras] {10.1093/mnras/sts383}, \href
  {https://ui.adsabs.harvard.edu/abs/2013MNRAS.429.1001A} {429, 1001}

\bibitem[\protect\citeauthoryear{{Carroll} \& {Strassmeier}}{{Carroll} \&
  {Strassmeier}}{2014}]{carroll2014}
{Carroll} T.~A.,  {Strassmeier} K.~G.,  2014, \mn@doi [\aap]
  {10.1051/0004-6361/201322825}, \href
  {http://adsabs.harvard.edu/abs/2014A\%26A...563A..56C} {563, A56}

\bibitem[\protect\citeauthoryear{{Donati} \& {Landstreet}}{{Donati} \&
  {Landstreet}}{2009}]{donati2009}
{Donati} J.-F.,  {Landstreet} J.~D.,  2009, \mn@doi [\araa]
  {10.1146/annurev-astro-082708-101833}, \href
  {http://adsabs.harvard.edu/abs/2009ARA\%26A..47..333D} {47, 333}

\bibitem[\protect\citeauthoryear{{Donati}, {Semel}, {Carter}, {Rees}  \&
  {Collier Cameron}}{{Donati} et~al.}{1997}]{donati1997}
{Donati} J.-F.,  {Semel} M.,  {Carter} B.~D.,  {Rees} D.~E.,   {Collier
  Cameron} A.,  1997, MNRAS, 291, 658

\bibitem[\protect\citeauthoryear{{Donati} et~al.,}{{Donati}
  et~al.}{2003}]{donati2003}
{Donati} J.~F.,  et~al., 2003, \mn@doi [\mnras]
  {10.1046/j.1365-2966.2003.07031.x}, \href
  {https://ui.adsabs.harvard.edu/abs/2003MNRAS.345.1145D} {345, 1145}

\bibitem[\protect\citeauthoryear{{Grunhut} et~al.,}{{Grunhut}
  et~al.}{2013}]{grunhut2013}
{Grunhut} J.~H.,  et~al., 2013, \mn@doi [\mnras] {10.1093/mnras/sts153}, \href
  {https://ui.adsabs.harvard.edu/abs/2013MNRAS.428.1686G} {428, 1686}

\bibitem[\protect\citeauthoryear{{Hill}, {Folsom}, {Donati}, {Herczeg},
  {Hussain}, {Alencar}, {Gregory}  \& {Matysse Collaboration}}{{Hill}
  et~al.}{2019}]{hill2019}
{Hill} C.~A.,  {Folsom} C.~P.,  {Donati} J.~F.,  {Herczeg} G.~J.,  {Hussain}
  G.~A.~J.,  {Alencar} S.~H.~P.,  {Gregory} S.~G.,   {Matysse Collaboration}
  2019, \mn@doi [\mnras] {10.1093/mnras/stz403}, \href
  {https://ui.adsabs.harvard.edu/abs/2019MNRAS.484.5810H} {484, 5810}

\bibitem[\protect\citeauthoryear{{Kochukhov}, {Makaganiuk}  \&
  {Piskunov}}{{Kochukhov} et~al.}{2010}]{kochu2010}
{Kochukhov} O.,  {Makaganiuk} V.,   {Piskunov} N.,  2010, \mn@doi [\aap]
  {10.1051/0004-6361/201015429}, \href
  {http://adsabs.harvard.edu/abs/2010A\%26A...524A...5K} {524, A5}

\bibitem[\protect\citeauthoryear{{Leone}, {Scalia}, {Gangi}, {Giarrusso},
  {Munari}, {Scuderi}, {Trigilio}  \& {Stift}}{{Leone}
  et~al.}{2017}]{leone2017}
{Leone} F.,  {Scalia} C.,  {Gangi} M.,  {Giarrusso} M.,  {Munari} M.,
  {Scuderi} S.,  {Trigilio} C.,   {Stift} M.~J.,  2017, \mn@doi [\apj]
  {10.3847/1538-4357/aa8d72}, \href
  {https://ui.adsabs.harvard.edu/abs/2017ApJ...848..107L} {848, 107}

\bibitem[\protect\citeauthoryear{{Marsden} et~al.,}{{Marsden}
  et~al.}{2014}]{marsden2014}
{Marsden} S.~C.,  et~al., 2014, \mn@doi [\mnras] {10.1093/mnras/stu1663}, \href
  {http://adsabs.harvard.edu/abs/2014MNRAS.444.3517M} {444, 3517}

\bibitem[\protect\citeauthoryear{{Mathys}}{{Mathys}}{1988}]{mathys1988}
{Mathys} G.,  1988, \aap, \href
  {https://ui.adsabs.harvard.edu/abs/1988A&A...189..179M} {189, 179}

\bibitem[\protect\citeauthoryear{{Mathys}}{{Mathys}}{1989}]{mathys1989}
{Mathys} G.,  1989, \fcp, \href
  {http://adsabs.harvard.edu/abs/1989FCPh...13..143M} {13, 143}

\bibitem[\protect\citeauthoryear{{Mathys}}{{Mathys}}{1991}]{mathys1991}
{Mathys} G.,  1991, \aaps, \href
  {https://ui.adsabs.harvard.edu/abs/1991A&AS...89..121M} {89, 121}

\bibitem[\protect\citeauthoryear{{Neiner}, {Alecian}, {Briquet}, {Floquet},
  {Fr{\'e}mat}, {Martayan}, {Thizy}  \& {Mimes Collaboration}}{{Neiner}
  et~al.}{2012}]{neiner2012}
{Neiner} C.,  {Alecian} E.,  {Briquet} M.,  {Floquet} M.,  {Fr{\'e}mat} Y.,
  {Martayan} C.,  {Thizy} O.,   {Mimes Collaboration} 2012, \mn@doi [\aap]
  {10.1051/0004-6361/201117941}, \href
  {https://ui.adsabs.harvard.edu/abs/2012A&A...537A.148N} {537, A148}

\bibitem[\protect\citeauthoryear{{Petit}, {Louge}, {Th{\'e}ado}, {Paletou},
  {Manset}, {Morin}, {Marsden}  \& {Jeffers}}{{Petit} et~al.}{2014}]{petit2014}
{Petit} P.,  {Louge} T.,  {Th{\'e}ado} S.,  {Paletou} F.,  {Manset} N.,
  {Morin} J.,  {Marsden} S.~C.,   {Jeffers} S.~V.,  2014, \mn@doi [\pasp]
  {10.1086/676976}, \href
  {https://ui.adsabs.harvard.edu/abs/2014PASP..126..469P} {126, 469}

\bibitem[\protect\citeauthoryear{{Ram{\'{\i}}rez V{\'e}lez}, {Y{\'a}{\~n}ez
  M{\'a}rquez}  \& {C{\'o}rdova Barbosa}}{{Ram{\'{\i}}rez V{\'e}lez}
  et~al.}{2018}]{jcrv2018}
{Ram{\'{\i}}rez V{\'e}lez} J.~C.,  {Y{\'a}{\~n}ez M{\'a}rquez} C.,
  {C{\'o}rdova Barbosa} J.~P.,  2018, \mn@doi [\aap]
  {10.1051/0004-6361/201833016}, \href
  {http://adsabs.harvard.edu/abs/2018A%26A...619A..22R} {619, A22}

\bibitem[\protect\citeauthoryear{{Rees} \& {Semel}}{{Rees} \&
  {Semel}}{1979}]{rees1979}
{Rees} D.~E.,  {Semel} M.~D.,  1979, \aap, \href
  {http://adsabs.harvard.edu/abs/1979A\%26A....74....1R} {74, 1}

\bibitem[\protect\citeauthoryear{{Ryabchikova}, {Piskunov}, {Kurucz},
  {Stempels}, {Heiter}, {Pakhomov}  \& {Barklem}}{{Ryabchikova}
  et~al.}{2015}]{vald2015}
{Ryabchikova} T.,  {Piskunov} N.,  {Kurucz} R.~L.,  {Stempels} H.~C.,  {Heiter}
  U.,  {Pakhomov} Y.,   {Barklem} P.~S.,  2015, \mn@doi [\physscr]
  {10.1088/0031-8949/90/5/054005}, \href
  {https://ui.adsabs.harvard.edu/abs/2015PhyS...90e4005R} {90, 054005}

\bibitem[\protect\citeauthoryear{{Sabin}, {Wade}  \& {L{\`e}bre}}{{Sabin}
  et~al.}{2015}]{sabin2015}
{Sabin} L.,  {Wade} G.~A.,   {L{\`e}bre} A.,  2015, \mn@doi [\mnras]
  {10.1093/mnras/stu2227}, \href
  {https://ui.adsabs.harvard.edu/abs/2015MNRAS.446.1988S} {446, 1988}

\bibitem[\protect\citeauthoryear{{Scalia}, {Leone}, {Gangi}, {Giarrusso}  \&
  {Stift}}{{Scalia} et~al.}{2017}]{scalia2017}
{Scalia} C.,  {Leone} F.,  {Gangi} M.,  {Giarrusso} M.,   {Stift} M.~J.,  2017,
  \mn@doi [\mnras] {10.1093/mnras/stx2090}, \href
  {https://ui.adsabs.harvard.edu/abs/2017MNRAS.472.3554S} {472, 3554}

\bibitem[\protect\citeauthoryear{{Semel}}{{Semel}}{1967}]{semel1967}
{Semel} M.,  1967, Annales d'Astrophysique, \href
  {https://ui.adsabs.harvard.edu/abs/1967AnAp...30..513S} {30, 513}

\bibitem[\protect\citeauthoryear{{Semel}}{{Semel}}{1995}]{semel1995}
{Semel} M.,  1995, in {Comte} G.,  {Marcelin} M.,  eds,  Astronomical Society
  of the Pacific Conference Series Vol. 71, IAU Colloq. 149: Tridimensional
  Optical Spectroscopic Methods in Astrophysics. p.~340

\bibitem[\protect\citeauthoryear{{Shorlin}, {Wade}, {Donati}, {Land street},
  {Petit}, {Sigut}  \& {Strasser}}{{Shorlin} et~al.}{2002}]{shorlin2002}
{Shorlin} S.~L.~S.,  {Wade} G.~A.,  {Donati} J.~F.,  {Land street} J.~D.,
  {Petit} P.,  {Sigut} T.~A.~A.,   {Strasser} S.,  2002, \mn@doi [\aap]
  {10.1051/0004-6361:20021192}, \href
  {https://ui.adsabs.harvard.edu/abs/2002A&A...392..637S} {392, 637}

\bibitem[\protect\citeauthoryear{{Silvester} et~al.,}{{Silvester}
  et~al.}{2009}]{silvester2009}
{Silvester} J.,  et~al., 2009, \mn@doi [\mnras]
  {10.1111/j.1365-2966.2009.15208.x}, \href
  {https://ui.adsabs.harvard.edu/abs/2009MNRAS.398.1505S} {398, 1505}

\bibitem[\protect\citeauthoryear{{Stibbs}}{{Stibbs}}{1950}]{stibbs1950}
{Stibbs} D.~W.~N.,  1950, \mn@doi [\mnras] {10.1093/mnras/110.4.395}, \href
  {https://ui.adsabs.harvard.edu/abs/1950MNRAS.110..395S} {110, 395}

\bibitem[\protect\citeauthoryear{{Stift}}{{Stift}}{1975}]{stift1975}
{Stift} M.~J.,  1975, \mn@doi [\mnras] {10.1093/mnras/172.1.133}, \href
  {http://adsabs.harvard.edu/abs/1975MNRAS.172..133S} {172, 133}

\bibitem[\protect\citeauthoryear{{Stift}}{{Stift}}{2000}]{stift2000}
{Stift} M.~J.,  2000, A Peculiar Newletter, 33, 27

\bibitem[\protect\citeauthoryear{{Valenti} \& {Fischer}}{{Valenti} \&
  {Fischer}}{2005}]{valenti2005}
{Valenti} J.~A.,  {Fischer} D.~A.,  2005, \mn@doi [\apjs] {10.1086/430500},
  \href {http://adsabs.harvard.edu/abs/2005ApJS..159..141V} {159, 141}

\bibitem[\protect\citeauthoryear{{Villebrun} et~al.,}{{Villebrun}
  et~al.}{2019}]{villebrun2019}
{Villebrun} F.,  et~al., 2019, \mn@doi [\aap] {10.1051/0004-6361/201833545},
  \href {https://ui.adsabs.harvard.edu/abs/2019A&A...622A..72V} {622, A72}

\bibitem[\protect\citeauthoryear{{Wade}, {Donati}, {Landstreet}  \&
  {Shorlin}}{{Wade} et~al.}{2000}]{wade2000}
{Wade} G.~A.,  {Donati} J.~F.,  {Landstreet} J.~D.,   {Shorlin} S.~L.~S.,
  2000, \mn@doi [\mnras] {10.1046/j.1365-8711.2000.03271.x}, \href
  {https://ui.adsabs.harvard.edu/abs/2000MNRAS.313..851W} {313, 851}

\makeatother
\end{thebibliography}

\end{document}